\documentstyle{mn}
\begin{document}
\title[F$\>$10214+4724]{IRAS F10214+4724: the inner 100pc}
\author[Lacy et al.]
{Mark Lacy$^{1}$, Steve Rawlings$^{1}$ \& Stephen Serjeant$^{2}$\\
$^{1}$Astrophysics, Department of Physics, Keble Road, Oxford, OX1 3RH\\
$^{2}$Astrophysics Group, Imperial College London, Blackett
Laboratory, Prince Consort Road, London SW7 2BZ} 
\maketitle
\begin{abstract}
We use new high resolution near-infrared spectroscopy
and our previously-published optical spectroscopy 
of the gravitationally-lensed Seyfert-2 galaxy F10214+4724 to study both
the links between the starburst and AGN in this object and the properties
of the line-emitting clouds in the inner narrow-line region.
Close inspection of the rest-frame UV spectrum reveals interstellar or
stellar absorption features consistent with a compact, moderately-reddened
starburst providing about half the UV light, and explaining the
dilution of the UV continuum polarisation relative to the broad emission
lines. Spectroscopy of the H$\alpha$/[N{\sc ii}]
line blend has enabled us to assess the relative contributions of the 
emission from the narrow-line region of the Seyfert-2,  
and a moderately-reddened emission line region which we argue is 
associated with the starburst activity. Estimates of the star
formation rate from the unpolarised 
UV continuum flux and the H$\alpha$ flux are 
consistent to within their associated uncertainties.
We find we can plausibly explain the unusual emission line 
properties of F10214+4724 in terms
of conventional models for nearby Seyfert-2 galaxies if lensing is  
preferentially magnifying the side of the inner narrow-line region
between the AGN and the observer, and the other side is both less
magnified and partially obscured by the torus. 
The hydrogen densities of clouds in this region are high enough
to make the Balmer lines optically thick and to suppress forbidden
emission lines with low critical densities. From the 
emission-line spectrum we have deduced the column density 
of both ionised and neutral gas in the narrow-line clouds, and the
density of the ionised gas. Using these we have been able
to estimate the mass of the inner narrow-line clouds to be 
$\sim 1 M_{\odot}$, and show that
the gas:dust ratio $N_{\rm H}/E(B-V)$ in these clouds must be 
$\sim 1.3 \times 10^{27}$m$^{-2}$mag$^{-1}$, significantly higher
than average value in the Milky Way, 
$\sim 4.5 \times 10^{25}$m$^{-2}$mag$^{-1}$. The column density and low dust
content of a typical cloud are consistent with the properties of 
the warm absorbers seen in the X-ray spectra of Seyfert-1 galaxies. 
Our results thus favour models in which the narrow-line clouds start life
close to the nucleus and flow out.
An emission line from the lensing system has allowed us  to confirm its 
redshift as $z\approx 0.9$. 
\end{abstract}

\section{Introduction}

The gravitationally-lensed Seyfert-2 galaxy FSC 10214+4724 has been 
the subject of extensive study in all accessible wavebands since its
discovery (Rowan-Robinson et al.\ 1991). Initially thought to be the most 
luminous galaxy in the Universe, it became increasingly clear that 
gravitational lensing was at least partly responsible for its ultra-high 
luminosity [Matthews et al.\ 1994; Trentham 1995; Serjeant et al.\
1995 (Paper I); 
Broadhurst \& Leh\'{a}r 1995; Graham \& Liu 1995; Eisenhardt et al.\ 1996]. In 
parallel with this, the debate continued about the relative contributions 
of the Seyfert-2 and starburst components to the spectral energy 
distribution, a debate 
which is still on-going [e.g.\ Rowan-Robinson et al.\ 1993; 
Elston et al.\ 1994; Lawrence et al.\ 1994;  
Soifer et al.\ 1995; 
Goodrich et al.\ 1996; Kroker et al.\ 1996; Green \& Rowan-Robinson
1996; Serjeant et al.\ 1998
(Paper II)]. 

This latter controversy has assumed particular importance
recently with the inference of an apparent peak in the star formation rate of
the Universe at $z\sim 2$ deduced from the UV fluxes of high redshift 
galaxies (Madau, Pozetti \& Dickinson 1996). 
Based on this, we might expect all massive galaxies at $z\sim 2$ 
to be associated with star formation rates about ten times greater than in 
the local Universe (i.e.\ a few solar
masses per year for a $\sim L_*$ galaxy), consistent with the value
obtained from the narrow H$\alpha$ emission in F10214+4724
(Serjeant et al.\ 1998). We do not, however,
necessarily expect to find the extremely high star formation rates that are
implied by the far-infrared luminosity of F10214+4724 
if a large fraction of it is due to a starburst. 
Nevertheless, the possibility that a large amount of
star formation in the high-redshift Universe is hidden by dust
(and therefore missed from UV-based estimates) is suggested
by the discovery of an infrared background (Guiderdoni et al.\ 1997)
and by results of infrared and submm surveys (Rowan-Robinson et al.\ 1997; 
Smail, Ivison \& Blain 1997).    
The nuclear regions of AGN with their dense, dusty neutral tori
wrapping the central object may turn out to be ideal sites for hiding 
substantial starbursts.

The gravitational lensing of F10214+4724 also offers us an
opportunity to examine the inner regions of a luminous narrow-line AGN at high
effective spatial resolution. The arc of F10214+4724 is believed to
consist of three merged images within a total arc length of 0.7
arcsec. The total magnification of the arc is $\approx 100$, 
and because the lensing is modelled well by an
isothermal potential (Eisenhardt et al.\ 1996), the magnification
of each image occurs predominately along the direction of the arc, with
little magnification or demagnification perpendicular to it. So if
F10214+4724 were not magnified, the nuclear region would have an angular
size of only $\sim 7$ maS, corresponding to a physical size of 50 pc for 
$H_0=50$ kms$^{-1}$Mpc$^{-1}$ and $q_0=0.5$ (assumed throughout this
paper). Resolution of AGN on this physical scale is otherwise 
only possible for the most nearby objects using the Hubble Space
Telescope (HST).

In Paper II we showed that the Ly$\alpha$ photons from F10214+4724
emerge from a
neutral column of $\approx 2.5 \times 10^{25}$m$^{-2}$, and that the 
1:1 doublet ratio of O{\sc vi} could be produced by absorption in the
damping wings of Ly$\beta$ as the emission lines propagated through
this column. We further argued
that the neutral column was probably within the narrow-line region itself,
located at the back of radiation-bounded narrow-line clouds. We also
presented the results of photoionisation modelling which pointed to
the narrow-line emission arising in relatively dense,
highly-ionisation narrow-line clouds. Our models [and those of Soifer
et al.\ (1995)] were, however, unsuccessful in fully accounting for
the weak Balmer line emission and the lack of [O{\sc ii}]372.7 in the
spectrum.

To attempt to resolve some of the remaining questions about F$\,$10214+4724, 
we obtained near-infrared spectra in $J$-,$H$- and $K$-band with the CGS4 
spectrometer 
on the United Kingdom Infrared Telescope (UKIRT). 
The probable discovery of an emission line from the lensing system is 
discussed in section 3. 
In section 4 we discuss the Balmer lines, and describe 
how the spectra were used to estimate the relative contributions of the 
narrower component of H$\alpha$, associated with star formation in
Paper II and the H$\alpha$ from the highly-magnified inner narrow-line 
region (INLR). We also discuss the related
problem of the anomalously low H$\beta$ flux measured by us in Paper
II and by others, and   
place a firm limit on the contribution of any broader 
component to the H$\alpha$ flux from the broad-line region of the 
hidden quasar. Section 5 discusses the other near-infrared lines,
and section 6 the reddening towards the narrow-line region. 
In section 7, features in the 
the rest-frame UV spectrum of F10214+4724 are discussed and  
amount of the star-formation activity in 
F$\,$10214+4724 is estimated.  Having established that the Balmer line flux
from the INLR is very low, in section 8 we attempt to explain the 
observed emission 
line ratios from F10214+4742 in terms of the conventional model for Seyfert-2
galaxies. Section 9 discusses the properties of the clouds in the
inner narrow-line region. Finally in section 10 our results are
summarised and some of their implications discussed.

\section{New Infrared Spectroscopy}

F10214+4724 was observed with CGS4 on the UKIRT on the nights of 
1996 April 16,17 and 18 in $H$- and $K$-band 
with the 150 line mm$^{-1}$ grating, and on 
1997 February 1 with the 75 line mm$^{-1}$ grating in $J$-band as part of the
UKIRT service programme. A two-pixel 
(2.4-arcsec) slit was used for all the observations. Spectra were taken as 
detailed in Table 1, and were nodded 
along the slit by 10 or 25 pixels in the ABBAABBA.. sequence described by 
Eales \& Rawlings (1993). This results in positive and negative object
spectra appearing on the array. Various position angles
were used in an attempt to detect emission lines from galaxies in the group
containing the lensing galaxy.

\begin{table*}
\caption{CGS4 observations of F10214+4724}
\begin{tabular}{ccrcrrr}
Date& Central wavelength &Grating &Grating order& PA & Integration time& Flux standard\\
    &/$\mu$m            &          &    &/deg & /minutes & and magnitude    \\
16/04/96 & 2.265 &150 l/mm& 1 & 90 & 64&HD 105601 ($K=6.69$) \\
16/04/96 & 2.265 &150 l/mm& 1 & 0  &128&HD 105601 ($K=6.69$)\\
17/04/96 & 1.580 &150 l/mm& 2 & 50 & 64&BS 4039 ($H=4.52$) \\
17/04/96 & 1.580 &150 l/mm& 2 & 22 & 64&BS 4039 ($H=4.52$) \\
18/04/96 & 2.140 &150 l/mm& 2 & 90 & 96&BS 4030 ($K=4.40$) \\ 
01/02/97 & 1.220 & 75 l/mm& 2 & 22 & 96&BS 4051 ($J=5.58$) \\

\end{tabular}
\end{table*}

The data were wavelength and flux calibrated in {\sc iraf} using argon or 
krypton
arc spectra and the flux calibration standards listed in the table. A 
3rd-order
polynomial fit to the columns was used to improve the subtraction of the sky 
lines. Additional 
spectra of bright stars of spectral types A, F and G were used to aid in the 
identification and removal of atmospheric absorption features.
One-dimensional spectra were then extracted 
from the positive and negative traces and averaged together.

\section{An emission line from the lensing system?}

An emission line is seen in the spectrum offset by about one pixel (1.2 arcsec)
from the object spectrum at 1.256$\mu$m (Fig.\ 1). 
This does not correspond to any 
known emission line at the redshift of F10214+4724, and we therefore 
assume it originates in the lensing system, probably in  source 2 
of Matthews et al.\ (1994), but with a possible contribution from source 3.
As both source 2 and 3 have similar spectral energy distributions in 
the optical, which we identified with galaxies at 
$z\approx 0.9$ in Paper I, 
we have identified the emission line as H$\alpha$ at 
$z=0.914$. This is slightly different from the absorption line redshift
of 0.893 tentatively obtained by Goodrich et al.\ (1996), but this may 
either be due to errors in identifying and measuring these weak features (see 
discussion in Section 7), or due to one 
redshift being that of source 2 and the other that of source 3. If so, 
this gives a velocity difference of 3300 kms$^{-1}$, large for a galaxy
group but possibly consistent with sources 2 and 3 being members of the
same galaxy cluster.

\begin{figure}
\setlength{\unitlength}{1mm}
\begin{picture}(75,100)
\put(-25,-93){\includegraphics{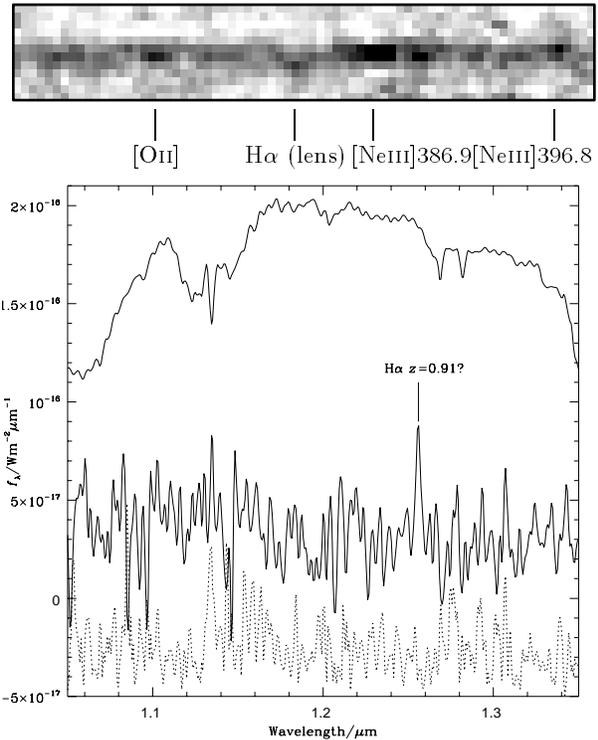}}
\end{picture}
\caption{{Top: greyscale of part of our $J$-band spectrum of
F10214+4724 showing the candidate emission line from the lens and the 
[O{\sc ii}] and [Ne{\sc iii}] emission lines from the object. Bottom: 
the $J$-band spectrum extracted $\approx 2$ arcsec along the slit 
showing the emission line from the lensing galaxy (lower solid curve). Lower
curve (dotted): noise estimate from the mean of three extractions of
sky near to the object, displaced by 
$-5.0\times 10^{-17}$ Wm$^{-2}\mu$m$^{-1}$. Upper curve: relative atmospheric 
transmission as a function of wavelength, multiplied by 
$2.0\times 10^{-16}$Wm$^{-2}\mu$m$^{-1}$ }}
\end{figure}

\section{The Balmer lines from F10214+4724}

Inspection of the H$\alpha$/[N{\sc ii}] blend at our highest resolution 
(FWHM of 0.0012 $\mu$m, or a resolving power of 1800) 
reveals a complicated structure, rendered more so by a bright sky 
line at 2.1518$\mu$m, the OH$^-$ $Q$ transition (Ramsay, Mountain \& 
Geballe 1992) which is saturated in our spectrum, and
is coincident with the [N{\sc ii}]654.8 line. 
Nevertheless, it seems clear that the [N{\sc ii}]658.4
line has a width consistent with the rest of the emission from the INLR, 
i.e.\ about 1000kms$^{-1}$, whereas the H$\alpha$ line has a component 
which is significantly narrower. We therefore attempted a $\chi^2$ fit 
to the data with eight free parameters: a constant continuum level, 
a wavelength, flux and width for the [N{\sc ii}]658.4 line (the properties 
of the [N{\sc ii}]654.8 line are then fixed as the ratio [N{\sc ii}]658.4/
[N{\sc ii}]654.8=3), a wavelength, flux and width for the narrow component
of H$\alpha$, and finally a flux for the component of H$\alpha$ which has the 
same width and redshift as the [N{\sc ii}] emission. The 
region of the saturated 
sky line was excluded from the fit, as was the part of the spectrum around 
2.178$\mu$m which was struck by a cosmic ray. 
The results of this fit are given 
in Table 2, and in Fig.\ 2 the results are plotted and compared to the data.

We found that small changes in e.g.\ the details of extraction and removal
of bad pixels resulted in a large range in the flux of the H$\alpha$ component
from the INLR (the flux of this component is probably only accurate to about a 
factor of three), 
but the basic results, namely that the [N{\sc ii}] lines are dominated by a 
component 
similar in width to the narrow lines in the optical spectrum, whereas the 
H$\alpha$ emission is dominated by a significantly narrower 
(FWHM $\approx 220$kms$^{-1}$)
component seem robust. The closeness of the FWHM of the narrow H$\alpha$ 
component to that of the CO 3-2 line (Radford et al.\ 1996) make it 
tempting to identify the narrow H$\alpha$ emission line with star-forming 
activity (cf.\ Paper II).

Comparison of our results with previous attempts at deblending the 
H$\alpha$/[N{\sc ii}] complex (Elston et al.\ 1994;
Kroker et al.\ 1996; Paper II) make it clear that the 
high spectral resolution is required to detect the narrower
H$\alpha$ component. In common with Elston et al.\ (1994) and Paper II
but at variance with Kroker et al.\ (1996) we find no evidence of an H$\alpha$
component from the broad line region. Comparing with Paper II, we find
that our higher resolution spectrum shows that the [N{\sc ii}]658.4
line is dominated by the INLR component, and the contribution of any
narrow component associated with the star-forming region 
to the [N{\sc ii}] flux
must be small. As a consequence, we find that the H$\alpha$ flux
from the INLR is much smaller than that estimated in Paper II, with
most of the flux in the H$\alpha$+[N{\sc ii}] complex coming from
the INLR component of [N{\sc ii}].

To obtain an upper limits on the fluxes of H$\alpha$ lines from the
broad line region (BLR), we have added 
in an artificial broad component to the first order $K$-band spectrum.
To begin with, we assumed a FWHM of 10000 kms$^{-1}$ for the 
broad line, based on the Keck spectrapolarimetry of Goodrich et al.\ (1996)
who measure this width for the broad component of the C{\sc iii}]190.9 line.
Our limit on the broad H$\alpha$ flux from this was obtained from the first
order $K$-band spectrum to be $\approx 5\times 10^{-18}$ Wm$^{-2}$, 
corresponding to a signal:noise of about $2\sigma$ per resolution element at 
half maximum intensity. We also tried a 4000 kms$^{-1}$ FWHM line, the broad 
C{\sc iii}] linewidth measured in direct light in Paper II. 
We have placed a similarly-derived limit of 
$\approx 2\times 10^{-18}$ Wm$^{-2}$ on the flux of such a component. 
This compares with the claimed detection of a broad line with a flux of 
$\approx 3.7\times 10^{-18}$ Wm$^{-2}$ and a width of 2400 kms$^{-1}$
by Kroker et al.\ (1996), which on the basis of these calculations we
should have easily detected.

\begin{table}
\caption{Results of a fit to the H$\alpha$/[N{\sc ii}] blend}
\begin{tabular}{lrrrr}
Line & wavelength & redshift & flux & FWHM \\
   & /$\mu$m     &          & /$10^{-19}$ & /kms$^{-1}$ \\
     &         &             &Wm$^{-2}$& \\
 $\,$ [N{\sc ii}]658.4&{\bf 2.1642} & 2.2871 & {\bf 28} & 
{\bf 1000} \\
$\,$ H$\alpha$ SF &{\bf 2.1576} & 2.2875 & {\bf 8.2}& {\bf 220} \\
$\,$ H$\alpha$ INLR & 2.1507 & 2.2871 & {\bf 1.0}& 1000 \\
\end{tabular}
Notes: items in bold face were free parameters in the model (see text).
\end{table}

\begin{figure}
\setlength{\unitlength}{1mm}
\begin{picture}(75,75)
\put(-5,-25){\includegraphics{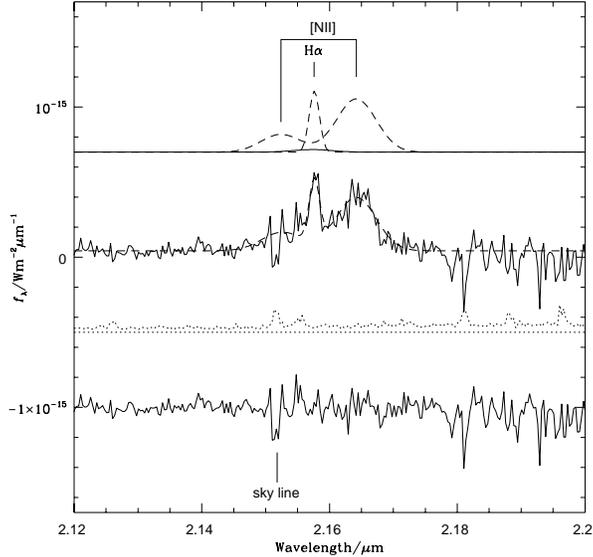}}
\end{picture}
\caption{The H$\alpha$/[N{\sc ii}]
blend and its fit. The top set of lines show the emission-line component 
fit to the blend. The [N{\sc ii}] and narrow H$\alpha$ components are shown 
dashed, the broader ``INLR'' H$\alpha$ component is the solid line. The next 
set of lines down are the data themselves, with the fit plotted through 
it as a dashed line, including the continuum level of 
$4.2\times 10^{-17} {\rm Wm^{-2}\mu m^{-1}}$ ($K\approx 17.5$). The dotted 
line below this is the noise as a function of wavelength that was used in the 
$\chi^2$ fit. The bottom solid line is the residual from the fit.
The position of a 
strong sky line near the weaker of the [N{\sc ii}] doublet is indicated. }
\end{figure}

\section{The $J$- and $H$-band spectra}

Fig.\ 3 shows the $J$-, $H$- and first order $K$-band spectra which between
them cover most of the available near-infrared window.
 
In the $J$-band, the [Ne{\sc v}]334.6 and 342.6 lines are clearly detected,
along with the [Ne{\sc iii}]386.9 line. Marginally detected are the 
[O{\sc ii}]372.7 and [Ne{\sc iii}]396.8 lines (see also Fig.\ 1). 
Interestingly, both our 
[O{\sc ii}] detection and that of Soifer et al.\ (1995) has the [O{\sc ii}]
emission redshifted with respect to the higher ionisation lines by about 
500 kms$^{-1}$.

The $H$-band spectrum shows a clearly detected and resolved 
[O{\sc iii}] doublet, and a marginal detection of He{\sc ii} 468.6. There
is no detection of H$\beta$ to a limit of $\approx 2\times 10^{-19}
{\rm Wm^{-2}}$. 

These spectra are consistent with earlier observations at lower 
spectral resolution (Soifer et al., 1995; Iwamuro et al.\ 1995; Paper II), 
and confirm the low [O{\sc ii}]372.7 and H$\beta$ fluxes found
by other groups. The higher resolution of these observations show that the 
FWHM of the lines is similar to those in the rest-frame UV, i.e.\ 
$\approx 1000$kms$^{-1}$.

\begin{table}
\caption{The $J$- and $H$-band spectra}
\begin{tabular}{lrrrr}
Line  & wavelength & redshift & flux & FWHM \\
      & /$\mu$m     &       & /$10^{-19}$& /kms$^{-1}$ \\
      &             &       &   Wm$^{-2}$ & \\
 $\,$ [Ne{\sc v}]334.6 &1.0982&2.2820&$7.1$& - \\ 
 $\,$ [Ne{\sc v}]342.6 &1.1247&2.2828&$18$&1500\\
 $\,$ [O{\sc ii}]372.7 &1.2301&2.3005&$2$& - \\
 $\,$ [Ne{\sc iii}]386.9&1.27237&2.2886&$13$&- \\
 $\,$ [Ne{\sc iii}]396.8&1.30419&2.2868&$1$& - \\
 $\,$ He{\sc ii} 468.6 &1.5397&2.2857&$2.3$&-\\
 $\,$ [O{\sc iii}]495.9&1.6292&2.2854&$18$& 810 \\
 $\,$ [O{\sc iii}]500.7&1.6450&2.2853&$53$ & 1040 \\
\end{tabular}
\end{table}

\begin{figure*}
\setlength{\unitlength}{1mm}
\begin{picture}(150,75)
\put(-25,-108){\includegraphics{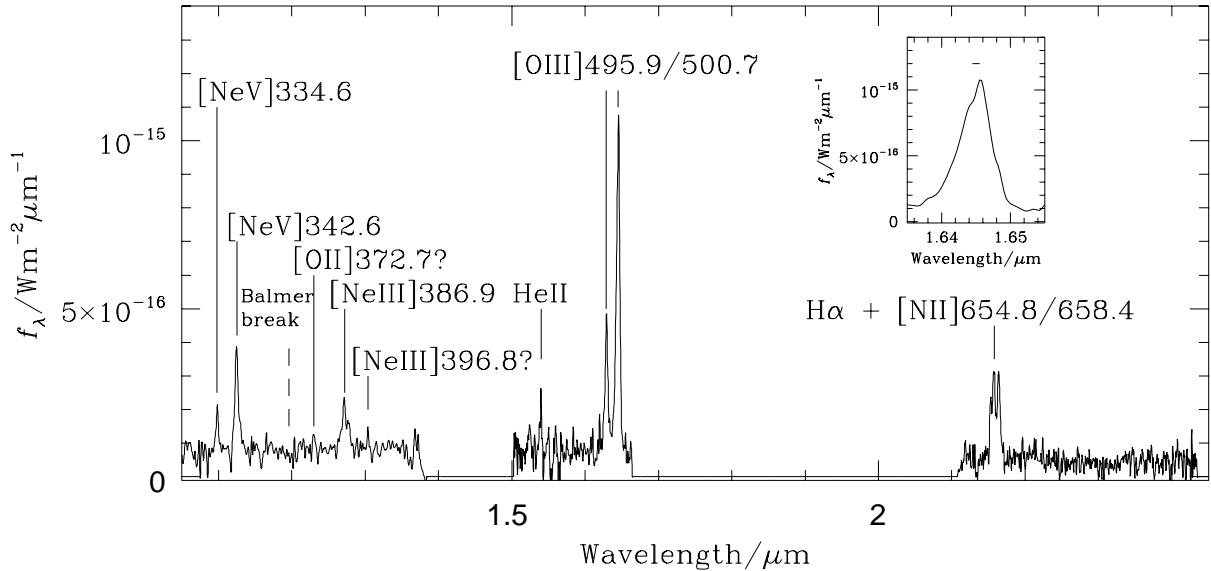}}
\end{picture}
\caption{ The near-infrared
spectrum of F10214+4724. The $J$- and $K$-band spectra have been 
smoothed with a three-pixel smoothing box, the $H$-band spectrum
(at higher resolution originally) has been smoothed twice. The
position of the Balmer break is indicated. Inset: detail of the [O{\sc
iii}]500.7 line showing the blue wing; the FWHM of the instrumental
profile is shown as a bar centred on the peak of the line.}
\end{figure*}

\section{The reddening towards the narrow-line region}

Our spectra all seem to be consistent with a low reddening towards the  
narrow-line region. Using the data from Paper II, the ratio of 
He{\sc ii} 468.6 / 164.0 is 4.3 (assuming our He{\sc ii} detection is real), 
compared to the case B value of $\approx 6.5$ (Osterbrock 1989). This
corresponds to an $A_{V}\approx 0.3$ if the reddening occurs in the source, 
much lower than the estimate of Soifer et al.\ (1995) ($A_{V}\approx 1.1$)
due to our lower He{\sc ii} 4686 flux measurement. Inspection of the 
spectrum of Soifer et al.\ (1995) suggests that the reason for this is 
a poor continuum level determination in their low-resolution spectrum.
Our reddening estimate is consistent with that derived from the 
UV He{\sc ii} 108.6: He{\sc ii} 164.0 line ratio in Paper II.

\begin{figure*}
\setlength{\unitlength}{1mm}
\begin{picture}(150,90)
\put(0,100){\includegraphics{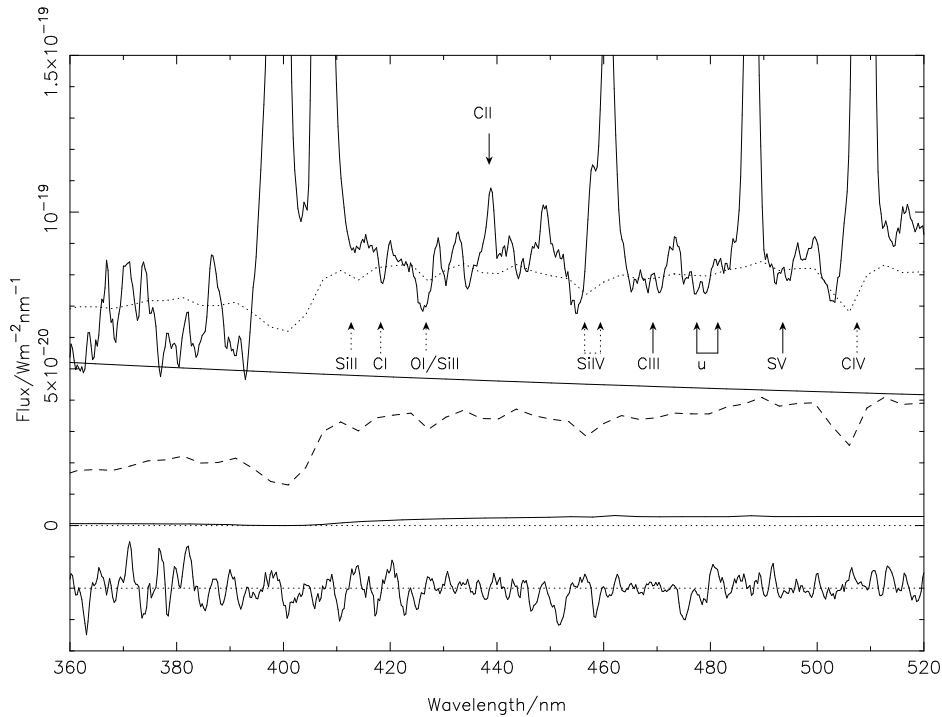}}
\end{picture}
\caption{ The UV spectrum of F10214+4724 from Paper II smoothed with a
5-pixel (1.4 nm) box-car filter. Positions of interstellar (dotted) and stellar
(solid) features are indicated by arrows. Note that the arrows for the
interstellar features are blueshifted by 1000 kms$^{-1}$ relative to
$z=2.286$. The smooth solid line below the spectrum is the power-law
polarised continuum of Goodrich et al.\ (1996) scaled by a factor of
two to allow for polarisation dilution. The dashed line is the
starburst spectrum and the solid line below it the nebular continuum 
contribution. The sum of these components is the dotted line plotted
through the spectrum. A noise spectrum is plotted at the bottom,
displaced by $-2.0\times 10^{-20}$}
\end{figure*}

\section{The FC2 component}

On the basis of their spectrapolarimetry, Goodrich et al.\ (1996) 
have argued for the presence of an unpolarised FC2 component in 
the UV spectrum of F10214+4724 with a redder slope than the polarised
emission and which contributes about half the UV flux. To investigate
the nature of this component we examined our WHT spectrum  
(Paper II) for UV absorption features. This spectrum is shown in Fig.\
4, scaled so as to show up the details of the continuum (the
emission lines are discussed in Paper II). 
Lists of interstellar and stellar absorption features from Verner,
Barthel \& Tytler (1994), Kinney et al.\ (1993) and Dey et al.\ (1997)
were used.

The strong interstellar line blend of O{\sc i} 130.2/Si{\sc ii} 130.4 
and the stellar wind 
or interstellar Si{\sc iv} 139.4/140.3 lines are 
convincingly detected. C{\sc ii} 133.5 is also seen in emission. 
All these features are also
visible in the Keck spectrum of Goodrich et al.\ (1996).

Less convincingly detected are interstellar lines of 
C{\sc i} 127.7 and the interstellar/stellar wind C{\sc iv} 154.8/155.1 lines 
in absorption. Finally, much less convincing are the interstellar  
Si{\sc ii} 126.0 line and some stellar features. These
stellar features are identifiable as such because they do not correspond to
transitions to or from the ground state. C{\sc iii} 142.8, an unidentified
pair of lines at 145.3 and 146.5, and S{\sc v} 150.2 are all very marginally
detected. The shape of the spectrum and position of the candidate
absorption features are, however, consistent with the spectrum of Goodrich et
al.\ (1996). We also note the possibility that the absorption
features around 530 nm noted by Goodrich et al.\ and attributed to 
Mg{\sc ii} at $z=0.892$  also correspond to a
blend of stellar absorption features at 160-163 nm in the 
linelist of Kinney et al.\
(1993). This may also help to account for the apparent discrepancy in
the emission and absorption-line redshifts for the lens.
The reality of these stellar absorption features is hard to establish,
as their signal-to-noise is
low, and the true continuum level hard to define due to the possibility
of confusion by unidentified emission lines associated with the AGN
activity. We
cannot thus claim to have found definitive evidence for a young
stellar population, but the similarity between this spectrum and those
of high redshift starburst galaxies identified through Lyman dropout
techniques (e.g.\ Lowenthal et al.\ 1997) is strong once the
mostly AGN-powered emission lines are subtracted.

The absorption troughs of the interstellar and stellar wind lines are
all blueshifted with respect to the mean emission line redshift of 2.286.
The deepest interstellar line, that of the O{\sc i}/Si{\sc ii} blend 
is blueshifted by $\approx 1200-1700$kms$^{-1}$ (depending on which of
the lines dominates the blend), indicative of a rapid outflow of low ionisation
gas. The Si{\sc iv} and C{\sc iv} stellar wind absorption troughs are 
even more highly 
blueshifted, by up to 4000 kms$^{-1}$, as expected if they arise in 
supergiant winds partly filled in by the AGN emission lines. 
In contrast, C{\sc ii} 133.5 is seen in
emission at the emission-line redshift, although there is a hint of
absorption in the blue wing, resulting in a P-Cygni profile. 
Blueshifting of the interstellar absorption lines is
seen in both low and high redshift starbursts (Heckman 1998), 
and it seems plausible that the blueshifts of the 
interstellar lines are due to fast winds or superwinds  
from the starburst region.

To estimate the star formation rate (SFR) required to produce the FC2
component, we have used a population
synthesis model from the PEGASE database of Fioc \& Rocca-Volmerange (1997)
and the starburst reddening law of
Calzetti et al.\ (1994). The attenuation was calculated from the
reddening law according to the prescription of 
Meurer et al.\ (1995). To estimate the contribution of the scattered
nuclear light (the FC1 component) we take the mean polarisation of the broad
emission lines of Goodrich et al.\ (1996) to be 50 per cent. We then assume
these are purely scattered light and correct the polarised
continuum by a factor of two to make a rough correction for
polarisation dilution of the FC1. We also assume that the star
formation rate has been roughly constant for $\stackrel{>}{_{\sim}}
10^7$yr so that the age of the starburst is longer than the lifetimes
of the stars dominating the UV continuum. The reddening can then be
estimated by requiring that the shape of the UV emission matches that observed.
In this ``steady state'' approximation we derive an SFR, corrected for
a reddening of $E(B-V)\approx 0.6$ towards the continuum, 
of $\approx 8000 M_{\odot}/m \, {\rm yr^{-1}}$ for a Scalo (1986) IMF or 
$\approx 4000M_{\odot}/m \, {\rm yr^{-1}}$ for a Salpeter (1955) IMF,
where $m$ is the magnification of the star-forming region. 

If we identify the kinematically distinct [and spatially distinct
according to Kroker et al.\ (1996)] narrow ($220$kms$^{-1}$) H$\alpha$ line
with star formation we can attempt
to place an independent constraint on the star formation rate in F10214+4724
from this. The reddening 
estimate here depends on the H$\alpha$:H$\beta$ flux ratio, and a convincing 
detection of H$\beta$ has yet to be made. Nevertheless, the lowest limit, 
that of 
Elston et al.\ (1994), and the lower error bar on the claimed detection 
of Iwamuro et al.\ (1995) are both consistent with a flux of 
$\approx 10^{-19}$Wm$^{-2}$. If we assume this value for the flux of the 
H$\beta$ line then the H$\alpha$:H$\beta$ ratio is about 9 (albeit with a 
high uncertainty due to the difficulty in accurately deblending the 
H$\alpha$+[N{\sc ii}] lines), corresponding to
a rest-frame $E(B-V)\approx 0.9$, and hence an extinction to H$\alpha$ of 
a factor of $\approx 7.5$, assuming 
a case B value of 2.87 for the intrinsic H$\alpha$/H$\beta$
ratio. Note that this is probably consistent with the lower value of
reddening found for the UV flux: Calzetti et al.\ (1994) find that 
the reddening seen towards the 
Balmer emission lines is typically greater that seen towards the
continuum in nearby starbursts, perhaps because the very massive 
stars producing the bulk of the ionising radiation are too short-lived
to escape from the dusty regions in which they are formed. With this
amount of reddening any [O{\sc ii}]372.7 emission from the starburst
would be undetectable in our spectrum.

The star formation rate from the H$\alpha$ luminosity in the starburst
component can be compared to the value predicted from the UV star formation
rate: for a Scalo IMF Gallego et al.\ (1995) obtain a conversion of
H$\alpha$ luminosity to star formation rate of 
$0.94 \times 10^{34}$ W$M_{\odot}^{-1}$yr, and for a Salpeter one 
Bunker (1996) obtains $3.108 \times 10^{34}$ W$M_{\odot}^{-1}$yr.
These give star formation rates of $2550$ and 740 $M_{\odot}/m \, {\rm
yr}^{-1}$ for Scalo and Salpeter IMFs respectively, corrected for reddening,
lower than those obtained from the UV. Given the large uncertainties
in the estimates of the star formation from the two different methods though,
their agreement to within an order of magnitude is probably as good as
can be expected.

It is clear from the HST $I$-band image of Eisenhardt et al.\
(1996) and from the $B$-band image of Broadhurst \& Lehar (1995) that
the UV continuum is dominated by the $\approx 0.7$-arcsec arc of highly
magnified emission. In contrast, about 30 per cent of the narrow
H$\alpha$ emission is seen to be extended over $\approx 2$ arcsec
according to Kroker et al.\ (1996). Thus it is plausible that some ionising
radiation escapes from the star-forming region to ionise gas further
out, which may itself have been expelled from the nucleus in a
``superwind'', or may be infalling; 
emission from this gas will less magnified (the magnification
scales approximately with the inverse of the 
size of the emission region). This gas could
of course equally well be ionised by the AGN, however. 

We can further compare these estimates to models for the far-infrared emission.
A limit on the size of any starburst region can be estimated from the
size of the radio source, which is approximately the same as that of
the bright UV source, as one would expect if a large fraction of the UV was 
from star formation. Green \& Rowan-Robinson (1996) show that, for a
starburst to fit within the radio source scale-size requires a high
optical depth ($\tau_{\rm UV}\approx 800$) model with several
starburst clouds along the line of sight. Such a model contributes 
$\approx 40$ per cent of the infrared luminosity, 
$\approx 2\times 10^{14}L_{\odot}$. Using the conversion of
infrared flux to SFR of Condon (1992) implies a SFR $\sim 20000M_{\odot}/m$
yr$^{-1}$ for stars $>5 M_{\odot}$, or a total of 
$\sim 10^5M_{\odot}/m$ yr$^{-1}$ assuming a Salpeter IMF. This is 1-2 orders
of magnitude higher than deduced from the UV and H$\alpha$ emission,
but the high optical depths necessary in the model to keep the source
size small would certainly imply that the bulk of the star formation
would be hidden. Further evidence that much of the starburst could be
hidden comes from Kroker et al.\ (1996) who suggest that the reddening
derived from H$\alpha$ could be an 
underestimate as the CO fluxes from F10214+4724
convert to $A_V\geq 100$ assuming galactic
gas:dust conversion factors.

The exact value of the magnification of the starburst region is controversial.
As Trentham (1995) shows, the probability of obtaining a high magnification 
is strongly dependent on the source size. He 
demonstrated that a magnification factor of $\gg 10$
was very unlikely if the source size was $\sim 1$ kpc or larger. 
Smaller sources could, however, be much more highly magnified. Other
arguments in favour of small magnifications ($m\sim 10$) come from the size
of the CO emission region, which is marginally resolved by 
Downes, Solomon \& Radford (1995) with a size of $1.5 \pm 0.4 \times
\leq 0.9$ arcsec$^{2}$, and modelling of the far-infrared flux (e.g.\ Green
\& Rowan-Robinson 1996). The larger size of the near-infrared arc (2 arcsec)
has also lead to suggestions that this corresponds to the dusty
starburst region, which would have an unmagnified size of $\approx 1$
kpc (Graham \& Liu 1995). 
The radio source and the UV source are, however, the same size (within
the errors) and this size ($0.7\times \leq 0.1$ arcsec$^2$), 
combined with the lens models, suggests higher values, 
$m \sim 50-100$ corresponding to source sizes of $\sim 40-80$ pc 
(Broadhurst \& Lehar 1995; Eisenhardt et al.\ 1996)
if the radio source is associated with the starburst. 

The minimum size for the infrared source based on black-body arguments 
provides another constraint. Using an effective
temperature of 140K, corresponding to the peak in the observed
mid-infrared flux, Eisenhardt et al.\ (1996) estimate
this minimum size to be 130 pc, corresponding to $m=42$. This small 
source size 
was assumed by Broadhurst \& Leh\'{a}r to prove an AGN origin for the
far-infrared emission, but compact starbursts may be common in
infrared--luminous galaxies. For example, 
the radio supernovae in the northeastern component of 
Arp 220 are contained within a
projected area of only $100 \times 200$ pc$^2$ (Smith et al.\ 1998).


There is nevertheless a factor of $\approx 2$ difference between the best
estimate of the UV magnification and the maximum IR magnification 
from the black-body argument. This could arise for
several reasons. First, the starburst may simply be smaller than the 
IR black-body size, giving a small radio/UV source with high
magnification. This is, however, statistically unlikely as it requires
the starburst to be well-centered on the cusp, even though, 
provided some part of the infared source lies on the
cusp, the infrared magnification will not depend strongly on the exact position
relative to the cusp. Alternatively, the starburst size may be comparable to  
the black-body size, but patchy extinction may lead to clumpy UV 
emission. If one of these clumps falls on the cusp it will be 
strongly magnified relative to the bulk of the infrared flux. The similar
lengths of the radio and UV arcs need not rule this out: if the
UV-emitting region is offset towards the edge of the infrared source 
the lengths of the UV and radio arcs could be similar, but the magnifications
different. Close to the caustic $m \propto b^{-1}$, where $b$ is the 
impact parameter (Broadhurst \& Leh\'{a}r 1995). Thus the length of
the arc $l \propto mr \approx$ constant if $b\approx r$, where 
$r$ is the size of the source. This hypothesis can be tested with a
detailed comparison of the HST and deep, high resolution radio images
which should show small differences in the structure of the arc
between radio and UV wavelengths. Lens modelling uncertainties and
uncertainty in the precise wavelength of the mid-infrared peak 
may also help to account for the difference in the magnifications.

To summarise: from our near-infrared studies we can place a
fairly conservative lower limit on the SFR in F10214+4724 of 
$\stackrel{>}{_{\sim}} 10 m_{100}^{-1} M_{\odot} {\rm yr}^{-1}$ where
$m_{100}$ is the magnification of the UV/optical arc in units of 100. 
This is consistent with the FC2 component being entirely produced by 
star formation, as is suggested by the similarity of the 
spectral features in the UV to those seen in starburst galaxies.
Our data is also consistent with much higher star formation rates if the bulk
of the star formation is hidden by dust, up to the $\sim 2.5\times 10^3
(m_{\rm FIR}/40)^{-1}M_{\odot}{\rm yr}^{-1}$ estimated from the 
far-infrared flux, where $m_{\rm FIR}$ is the magnification of the 
far-infrared flux.
The unobscured starburst region is compact, with a scale size $\sim 100$pc. 
High velocity winds are present in the interstellar gas. 
It is therefore likely that, as suggested in the case of
Mrk477 by Heckman et al. (1997), the star formation is taking place in the
outer regions of the obscuring torus, where conditions in the gas
(which is sufficiently distant from the AGN to be shielded from its
radiation field by the inner torus) are likely to support star formation.

\begin{table}
\caption{A comparison of the emission line ratios of F10214+4724
and NGC 1068}
\begin{tabular}{lrrrl}
Line & F10214+4724         &NCG1068  & F10214+4724    \\  
     & (raw)               & (raw)   & dereddened     \\
$\,$O{\sc vi} 103.2& 0.45  &0.76 &0.87  \\
$\,$O{\sc vi} 103.8& 0.43  &0.99 &0.82  \\
$\,$He{\sc ii} 108.5& 0.14 &0.17 &0.23  \\
$\,$C{\sc iii} 117.6& 0.04 &$\stackrel{<}{_{\sim}}0.07$&0.06  \\
$\,$Ly$\alpha$      & 1.0  &4.8 & 1.3  \\
$\,$N{\sc v} 124.0  & 1.4  &1.3 & 1.8  \\
$\,$C{\sc ii} 133.5 & 0.03 &0.16&0.03  \\
$\,$Si{\sc iv} 139.4& 0.40 &0.41&0.44  \\
$\,$+O{\sc iv} 140.3&      &    &      \\
$\,$N{\sc iv} 148.7 & 0.40 &0.24&0.42  \\
$\,$C{\sc iv} 154.9 & 1.9  &1.9 & 1.9  \\
$\,$[Ne{\sc iv}] 160.2&0.11&$\stackrel{<}{_{\sim}}0.07$&0.11   \\
$\,$He{\sc ii} 164.0 & 1   &  1  & 1 \\
$\,$[O{\sc iii}]166.5& $\stackrel{<}{_{\sim}}0.05$&$\stackrel{<}{_{\sim}}0.07$&
$\stackrel{<}{_{\sim}}$0.05\\
$\,$N{\sc iii}]175.0 & 0.20 &0.27  & 0.20\\
$\,$Mg{\sc vi} 180.6 & 0.15 &0.09  & 0.15\\
$\,$C{\sc iii}]190.9 & 0.65 &1.12  & 0.65\\
$\,$[Na{\sc v}] 206.7& 0.12 &0.05  & 0.14\\
$\,$N{\sc ii}]214.3  & 0.08 &0.07  & 0.10\\
$\,$C{\sc ii}]232.6  & 0.30 &0.25  & 0.32\\
$\,$[Ne{\sc iv}]242.2&0.52  &0.40  & 0.51\\
$\,$[Ne{\sc v}]342.6 & 1.50 &0.54  & 1.16\\
$\,$[O{\sc ii}]372.7 & 0.09 &0.57  & 0.07\\
$\,$[Ne{\sc iii}]386.9&1.13 &0.70  & 0.84\\
$\,$He{\sc ii} 468.6 & 0.23 &0.29  & 0.16\\
$\,$H$\beta$         &$\stackrel{<}{_{\sim}}$0.1&0.75&
$\stackrel{<}{_{\sim}}$0.07\\
$\,$[O{\sc iii}]500.7& 5.0 & 9.9&3.4  \\
$\,$[O{\sc i}]630.0& 0.18 & 0.46 & 0.11\\
$\,$H$\alpha$ & $\sim$0.1 & 3.35&$\sim$ 0.06\\
$\,$[N{\sc ii}]658.4 & 2.8 & 6.0 & 1.7\\
$\,$[S{\sc ii}]671.7/3.2& 0.17&1.10 & 0.10\\
$\,$[Ar{\sc iii}]713.6&0.12 & 0.33 & 0.07\\
$\,$[O{\sc ii}]732.5& 0.23 & 0.26 & 0.14\\
\end{tabular}

\vspace*{2.0mm}
Note: Line fluxes are normalised to He{\sc ii} 164.0. 
The reddening assumed for F10214+4724 is $E(B-V)=0.1$, and 
a standard galactic reddening curve 
(Cardelli, Clayton \& Mathis 1989) assumed. Emission line fluxes
for NGC1068 are from Kriss et al.\ (1992); Antonucci, Hurt 
\& Miller (1994); Snijders, Netzer \& Boksenberg
(1986);  Koski (1978), and Osterbrock \& Fulbright (1996). 
\end{table}

\section{The emission-line ratios}

The unusually weak [O{\sc ii}]372.7 and Balmer lines have been a particular
problem in explaining the emission-line properties of F10214+4724.
Rather than attempt to model the emission line ratios using a photoionisation
code, as attempted by Soifer et al.\ (1995) and by ourselves in Paper II, in 
this paper we 
have adopted a more empirical approach. Table 4 compares the 
emission line ratios of F10214+4724 with those of the nearby Seyfert-2 galaxy
NGC1068, normalised to He{\sc ii} 164.0. 
As can be seen, the line ratios are remarkably similar, including the 
He{\sc ii} reddening diagnostic lines (108.5, 164.0 and 468.6 nm) 
even with no attempt at dereddening the spectra [which in the case of 
NGC1068 at least
is complicated by apparently wavelength-dependent extinction corrections, 
(Ferguson, Ferland \& Pradhan 1995) and/or possible blending of the 
He{\sc ii} 108.5 line (Netzer 1997)]. The strong UV lines
(e.g.\ O{\sc vi}, N{\sc v}, C{\sc iv}, C{\sc iii}]) are all well-reproduced. 
Many of the optical line ratios are also very similar to within
a factor of two (e.g.\ [Ne{\sc iii}], [O{\sc iii}]500.7) with the
exception of the hydrogen lines, and lines having critical densities 
$n_{\rm crit}\stackrel{<}{_{\sim}} 10^{10} {\rm m^{-3}}$ ([O{\sc ii}]372.7,
[S{\sc ii}]671.7/673.1) which are suppressed by factors $\approx 6$. 
Lines with $n_{\rm crit} \sim 10^{11-12} {\rm m^{-3}}$ 
([N{\sc ii}]654.8/658.4, [O{\sc iii}]495.9/500.7) are low by factors 
$\approx 2$, but not very discrepant given the typical range in line
ratios (Simpson \& Ward 1996) and the uncertainty in the differential
reddening between the two objects. [Ne{\sc v}]342.6 is higher in F10214+4724 
by a factor of $\approx 3$, but there is a general trend for all the Ne lines
to be brighter than their counterparts in NGC1068, and so it may be at least
partly accounted for by an enhanced Ne abundance relative to NGC1068.

On this basis 
it seems likely that the low $n_{\rm crit}$ lines are being suppressed
because the nebular density in a typical cloud, 
$n \sim 10^{10-11} {\rm m^{-3}}$, probably because 
lensing is preferentially magnifying the INLR where the cloud densities 
are higher. In the Simpson \& Ward (1996) model, the 
cloud densities are $\propto r^{-2}$ where $r$ is the distance from the 
nucleus. The ionisation parameter is therefore independent of the distance 
from the nucleus, which helps to explain why magnification of the INLR
has not produced a much higher ionisation spectrum than that for normal 
luminous Seyfert-2s. 

This interpretation is supported by the HST spectrum
of NGC1068 taken by Caganoff et al.\ (1991) using a 0.3-arcsec
($\approx 30$ pc)
aperture centred on the UV continuum peak. Although few line fluxes
are given in the paper, we have attempted to measure line ratios from
their plot. Within this small aperture, the degree of
ionisation as measured by the [Ne{\sc v}]342.6:[Ne{\sc iii}]368.9
ratio seems higher that in our spectrum of F10214+4724 (1.8 compared
to 1.3) and also higher than in the wide aperture 
($2.7 \times 4.0$ arcsec$^2$) spectrum
of NGC1068 in Table 4 (0.8), which contains a factor of 9.4 more
H$\beta$ flux. This suggests a departure from the $n \propto
r^{-2}$ law, in the sense that the density is decreasing more slowly 
with radius than $r^{-2}$. Nevertheless, a change in the 
[Ne{\sc v}]342.6:[Ne{\sc iii}]368.9 ratio of only a factor of 2.3 in an
order of magnitude difference in aperture size suggests that the
approximation of constant $U$ with radius is probably not too bad. 

The Caganoff et al.\  spectrum has 
weaker [O{\sc ii}]372.7 and [S{\sc ii}]671.6/673.2 emission relative to
He{\sc ii} 468.6 than for the NGC1068 spectrum through the larger
aperture by factors of $\approx 2$ and $4$ respectively, and
also a weaker H$\beta$ line by a factor of $\approx 2$.
From the [O{\sc ii}] doublet ratio Caganoff et al.\ estimate $n\approx
1.2 \times 10^9$m$^{-3}$, about an order of magnitude less than we
estimate for F10214+4724, and this seems consistent with a
less dramatic suppression of the low $n_{\rm crit}$ and Balmer lines 
in the INLR of NGC1068 than in that of F10214+4724, despite the
higher ionisation of the small aperture NGC1068 spectrum.

We can probably rule out the alternative 
possibility that the lower ionisation lines are 
suppressed due to their being formed at the edge of the narrow-line clouds 
and therefore not being lensed by as large a factor. 
As discussed in section 7, magnification near a caustic $\sim b^{-1}$. 
The INLR is $\stackrel{>}{_{\sim}} 10$pc
from the nucleus, very large compared with the 
thickness of the ionised layer, only $\sim 10^{11}$m for a density 
$n \sim 10^{10.5}$m$^{-3}$. Thus differential magnification of the ionisation
structure of individual clouds is 
unlikely to affect the observed line ratios.

The low levels of the H lines present a bigger problem. In Paper II we
discussed the possibility of suppressing Ly$\alpha$ through resonant
scattering in an H{\sc i} column, and used the velocity splitting of
the Ly$\alpha$ components to estimate the column density. 
In this paper we have further shown that both the H$\alpha$ and H$\beta$
fluxes from the 
INLR are very weak, resolving the problem of having discrepant reddening 
estimates from the H and He lines, but we still need to explain the general 
weakness of the Balmer lines. We now consider three
possibilities to help explain the lack of Balmer lines:
(i) the Balmer lines are optically thick, (ii) the Balmer 
lines are weakened by underlying absorption, and (iii)
the INLR has a very high metallicity. Possibility (ii) can probably be 
completely discounted as we see no Balmer break in either 
our $J$-band spectrum or that of Soifer et al.\ (1995). 
Possibility (iii) can probably also be ruled out as we would not expect to 
see the He{\sc ii}:metal emission line ratios to be so similar to those in 
NGC1068, which has much brighter H lines. This leaves
optical thickness of the Balmer lines as the only remaining 
explanation. As we now show, it seems plausible as the Balmer line
optical depth is indeed significant.
Davidson \& Netzer (1979) show that the optical depth to H$\alpha$ is given by
\[ \tau_{{\rm H}\alpha} \approx \sigma_{{\rm H}\alpha} 
A_{{\rm Ly}\alpha}^{-1} \zeta F_{\rm H} K_1 \tau_{{\rm Ly}\alpha}, \]
where $\sigma_{{\rm H}\alpha}=10^{-16.4} {\rm m^{2}}$ at the line centre 
is the cross-section of the $n=2$ state to an 
H$\alpha$ photon, $\zeta$ and $K_1$ are factors of order unity, 
$F_{\rm H}$ is the flux of ionising photons, $\tau_{{\rm Ly}\alpha}$ is the 
optical depth to the Ly$\alpha$ line and  
$A_{{\rm Ly}\alpha} = 4.7\times 10^8 {\rm s^{-1}}$ is the 
Einstein $A$-coefficient of the $2p$ state. 
Assuming  a dimensionless ionisation parameter 
$U=0.1$ and a hydrogen density $n=10^{10.5}$m$^{-3}$ gives 
$F=cnU=9\times 10^{17} {\rm m^{-2}s^{-1}}$, so 
$\tau_{{\rm H}\alpha} \sim 10^{-7} \tau_{{\rm Ly}\alpha}$. 

According to Paper II the minimum column density in H{\sc i} 
required to produce the observed Ly$\alpha$ profile is 
$2.5 \times 10^{25} {\rm m^2}$, which, assuming a cross section to 
the line centre of Ly$\alpha$ of $4.5 \times 10^{-18} {\rm m^2}$ gives
$\tau_{{\rm Ly}\alpha}\stackrel{>}{_{\sim}} 1\times 10^8$ to the line centre. 
Thus $\tau_{{\rm H}\alpha} \stackrel{>}{_{\sim}} 10$. 

As Ferland \& Netzer  
(1979) point out, however, this estimate of the optical depth needs to 
be interpreted carefully. Most of the optical depth arises in the partially 
ionised region of the nebula, where the Ly$\alpha$ photons are being 
resonantly scattered and the population of the $n=2$ level is therefore
highest. Provided the cloud is radiation-bounded, escape of H$\alpha$ photons
from the illuminated side of the nebula is far more likely than escape from 
the non-illuminated side (as is also the case for Ly$\alpha$ photons). Thus
if we are observing the clouds from the non-illuminated side, we expect
the Balmer and Ly$\alpha$ lines to be substantially suppressed. This case is
therefore different to that for the UV resonance lines, e.g.\ O{\sc vi} and 
C{\sc iv}, which may also be optically thick in F1024+4724
(Paper II). In the cases of these lines there are no reflecting 
layers to backscatter the photons: provided they are not absorbed by dust 
they will eventually escape isotropically.

The lack of 
hydrogen lines in F10214+4724 can thus be explained as a geometrical effect.
Backscattering of the Balmer and Ly$\alpha$ lines in the INLR suppresses
them relative to the outer narrow-line region, 
where escape is easier. This effect will be most pronounced 
if the side of the INLR which is closest to the cusp in the lensing
potential is between us and the AGN (Fig.\ 5). In this geometry, we 
see the back sides of the most highly-magnified narrow-line 
clouds, through which 
little flux from H lines can escape. Heisler, Lumsden \& Bailey (1996)
show that polarised broad lines are only detected in Seyfert-2 galaxies whose 
torus axes are inclined relatively close to the angle at which the BLR
becomes visible, thought to be about 45 deg. Inclination at such 
an angle will also improve suppression of the H lines in our backscattering 
model. 

In a normal Seyfert-2
such as NGC1068, we see brighter H lines because the outer 
narrow-line region 
is optically thin to H$\alpha$, thus enabling us to see the Balmer lines 
from the nearside, and the contribution from the far side of the central object
is larger, allowing both Ly$\alpha$ and Balmer lines to be seen from there. 
It is also likely that apart from lensing, obscuration by the dusty 
molecular torus of the other side of the INLR helps to reduce the 
Balmer and Ly$\alpha$ contributions from the far side of the central object.
Partial obscuration of the INLR has been used to explain the differences
in [O{\sc iii}] emission line strengths between quasars and radio galaxies
(Hes, Barthel \& Fosbury 1993).

\begin{figure}
\setlength{\unitlength}{1mm}
\begin{picture}(75,65)
\put(0,60){\includegraphics{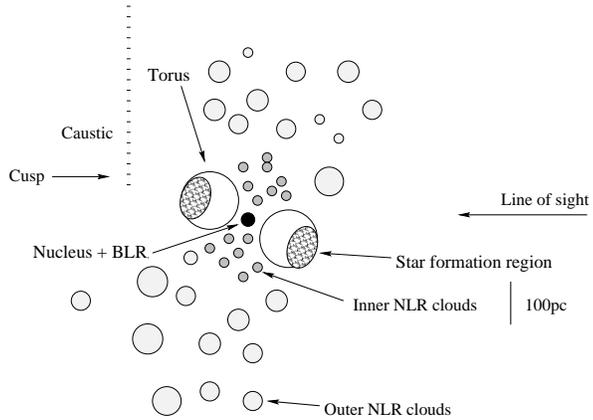}}
\end{picture}
\caption{ The geometry required to produce the observed emission line
properties of F10214+4724. The line marked ``caustic'' indicates the line of 
the caustic in the lensing potential, which reaches a cusp close to the 
nucleus and just above it [see fig.\ 1 of Broadhurst \& Leh\'{a}r (1995)], 
thus preferentially 
magnifying the forward-pointing side of the ionisation cone.}
\end{figure}

\section{Inferred properties of the inner narrow-line clouds}

Given an estimate of the density, ionisation parameter and the neutral
column, we can attempt to deduce the typical sizes and masses of the 
narrow-line clouds. The ionised column $N_{\rm H^+}$ can be estimated from the 
ionisation parameter, density and ionising flux from
\[ N_{\rm H^+} = \frac{Uc}{\alpha_{\rm B}}\approx 1 \times 10^{26} 
{\rm m^{-2}} \]
where $\alpha_{\rm B} = 2.58 \times 10^{-19} {\rm m^3 s^{-1}}$ is the 
case B recombination coefficient. To this, the neutral column adds a comparable
$\approx 2.5 \times 10^{25}{\rm m}^{-2}$ (Paper II).
Hence assuming a uniform density cloud, a typical size 
is $\sim 0.1$pc and a typical mass is $\sim 1M_{\odot}$, a 
little larger than the estimates in Paper II due to our inclusion of
the ionised column. 

We can compare our results to the theoretical predictions of Mathews \&
Veilleux (1989) who deduced from a stability analysis of the clouds in the 
narrow line region an $N_{\rm H} \sim 10^{26}-10^{28}{\rm m^{-2}}$. 
Clouds in outflowing winds with column densities in this regime are stable to 
Rayleigh-Taylor instabilities when radiative pressure and dynamical pressure 
act together, yet are small enough both to ensure a covering factor $<<1$ and 
smooth line profiles.

Extinction by dust in the inner narrow-line clouds would be expected to affect
the observed line ratios, particularly if, as our model predicts, we
are seeing emission-line light emerging through large column densities
in the narrow line clouds. From our derived column density (ionised
plus neutral)
and assuming a gas:dust ratio of $N_{\rm H}/E(B-V)= 4.5 \times 10^{25} {\rm
m^{-2}mag^{-1}}$ 
(Bohlin, Savage \& Drake 1978), we obtain $E(B-V) \approx 2.8$ mag,
much larger that the $\approx 0.1$ we measure (which of course is an upper
limit to the amount of reddening within the narrow-line clouds, as 
it does not include dust outside the nuclear region). The double 
Ly$\alpha$ line is also good evidence against dust in the neutral 
column, as even a small amount of dust would destroy Ly$\alpha$ 
photons. Thus the narrow-line
clouds must contain relatively low amounts of dust. Even at 10 pc
from the AGN, we are well away from the dust sublimation point,
thought to be just outside the broad-line region (Laor \& Draine
1993). The narrow-line clouds may start their lives close to the
AGN and then flow out, or perhaps have their dust destroyed by shocks
associated with the AGN.

Our column depths are also consistent with models in which the 
fully-ionised ``warm absorbers'' seen in
the X-ray spectra of Seyfert 1 galaxies and quasars, and possibly
related to associated absorbers seen in quasar spectra, are narrow-line
clouds seen close to the nucleus (e.g.\ George et al.\ 1998). These
have typical column densities $\sim 10^{25}-10^{27}{\rm m^{-2}}$, but
seem to be mostly dust free, and this, coupled with their short 
variability timescales, suggests an origin at least as close to the AGN
as the outer broad-line region. Outflow of these clouds into the 
narrow-line region in a decelerating flow (Crenshaw 1997) would
account for the lack of dust in the inner narrow-line clouds. 

\subsection{X-ray emission}

In common with most Seyfert 2 galaxies, the X-ray emission from F10214+4724
is weak (Lawrence et al.\ 1994), implying an absorbing H{\sc i} column
of $\sim 5\times 10^{27}{\rm m^{-2}}$, an order of magnitude higher
than that in out narrow line clouds, and consistent with the absorption
taking place in a torus of neutral gas around the AGN. Such a column 
implies a reddening $E(B-V)\sim 100$ if the gas:dust ratio is galactic, 
or $\stackrel{<}{_{\sim}} 5$ if it is the same as deduced from the
narrow line clouds. Either
would also be more than sufficient to hide the nucleus and broad-line region
in the UV and optical.

\section{Discussion}

Our studies of IRAS F10214+4724 have allowed us
to plausibly resolve many of the remaining puzzles relating to this 
object. Furthermore, the lensed nature of the galaxy has allowed us to
examine the nuclear region in detail only achievable with the HST on
the most nearby Seyfert galaxies. With this we have been able to argue
that there is a probably an intimate link between a starburst and AGN in this 
object, namely that the starburst is occuring very close in to the
nucleus, possibly in the obscuring torus. From optical depth
arguments we have been able to deduce many of the properties of the
narrow-line clouds in the INLR, and thus address some of the
issues associated with absorption of AGN spectra.

Fig.\ 5 shows the model we have chosen to adopt 
for the nuclear regions of F10214+4724. This gives the greatest suppression of
the hydrogen lines, although in fact any geometry in which the cusp falls
in front of the INLR will lead to some suppression.
The axis of the dusty molecular torus
is inclined such that the cusp of the lensing potential falls on the side
of the ionisation cone which points towards the line of sight. This enables 
us to scatter the Lyman and Balmer emission from the backside of the 
clouds in the highly-magnified INLR. The INLR has a similar scale
size ($\sim 50-100$pc) as the mid-infrared 
source needs to have if the intrinsic 
luminosity is to be comparable with local ultra-luminous IRAS galaxies
(Broadhurst \& Leh\'{a}r 1995; Eisenhardt et al.\ 1996), 
i.e.\ if the magnification $\sim 50-100$. The 
low density outer part of the narrow-line region 
has a much lower magnification, and the 
other side of the INLR both has a lower magnification and may be partially
obscured from our line of sight by the torus. Thus lines from the low density
clouds in the outer narrow-line region 
and the Balmer and Ly$\alpha$ lines appear anomalously weak.

\subsection{The nature of the inner narrow-line clouds}

Using the narrow-line spectrum, we can specify the properties of a 
``typical'' inner narrow-line cloud. Densities inferred from the
suppression of emission lines of low $n_{\rm crit}$ are $n \sim 10^{10.5}$
m$^{-3}$, with an ionisation parameter of $U \sim 0.1$ inferred from
the relative strength of high-ionisation lines. Column densities in ionised and
neutral gas are predicted to be comparable if the 
double-peaked Ly$\alpha$ line is
formed by propagation through the neutral backs of the narrow-line clouds.
This is consistent with the appearance of emission from neutral oxygen
in the spectrum, but requires the narrow-line clouds to have a low
dust content.

Our column density estimates are consistent with the narrow-line
clouds being in outflow from the nucleus, stabilised according to the 
mechanism of Mathews \& Vielleux (1989), with an origin close to the 
BLR to account for their low dust content. When they are closest to
the nucleus they may appear as warm absorbers in X-ray spectra.

Narrow-line profiles in AGN (including F10214+4724; see Fig.\ 3)
almost invariably have blue wings
and it has been recognised for some time that a
possible explanation of this is that the clouds are outflowing, with
clouds on the far side of the nucleus from the observer being
partially obscured (e.g.\ Osterbrock 1989). The alternative, that the
narrow-line clouds are infalling and dust absorption in the rear of
the clouds preferentially attenuates light from the nearside of the 
narrow-line region can, in the case of F10214+4724 at least, be ruled out.

\subsection{F10214+4724 in context}

The angular magnification
provided by gravitational lensing of F10214+4724 has allowed us to
observe the emission from
within the inner 100pc of the AGN, scales which can usually only be
studied in the most nearby objects.  
If our model for F10214+4724 is correct, the unlensed Seyfert-2 galaxy is 
remarkably similar to the most luminous local examples. The detection of 
polarised broad lines, an associated compact starburst and nuclear
radio emission all have precedents at low redshift. 

F10214+4724 is seen at a redshift close to the apparent peak in star
formation activity in the Universe, and has a UV component very
reminiscent of the Lyman break galaxies (e.g.\ Lowenthal et al.\ 1997),
although much more reddened. Once corrected for lens magnification
the star formation rate deduced from the H$\alpha$ and UV fluxes is
only a factor of a few to ten higher than predicted in the average massive 
galaxy at these epochs, although in the case of F10214+4724 it is
confined to a very small region near to the nucleus. The evidence for
dust both from the H$\alpha$/H$\beta$
line ratio and the slope of the UV spectrum is strong though, and it
is quite plausible that much of the evidence for star formation could be
hidden in the UV and optical, appearing only as an infrared
excess. Thus although the optical/UV evidence for a starburst in
F10214+4724 is fairly convincing, on the basis of our observations
alone we cannot tell which of the starburst or the AGN dominates the
mid-infrared flux. 

Surveys for objects like F10214+4724 are already succeeding. The ISO
study of the Hubble Deep Field (Rowan-Robinson et al.\ 1997) has found
several $z\stackrel{<}{_{\sim}}1$ galaxies with mid-infrared 
excesses indicative of dust shrouded starbursts, and  
recently, Ivison et al.\ (1998) have announced the first detection of 
a high-$z$ galaxy discovered with the SCUBA submm instrument, 
SMM 02399-0136, discovered as part of a survey behind the galaxy
cluster A370 (Smail et al.\ 1997). Like F10214+4724,
this galaxy is a narrow-line AGN with relatively weak Ly$\alpha$
emission and  blueshifted UV absorption
features, including C{\sc iv} and Si{\sc iv}, and possibly Si{\sc
ii}152.7 (although this looks more likely to be part of a broad 
absorption line trough from C{\sc iv}). However, SMM 02399-0136
has a much stronger broad C{\sc iii}]1909 component and no 
splitting of the Ly$\alpha$ line. Also, the
best estimates of the effects lensing for both objects make 
SMM 02399-0136 significantly more luminous than F10214+4724. Apart
from  possibly Si{\sc ii} there are no other low ionisation UV features, and
this, combined with the larger broad C{\sc iii}] component suggests that 
SMM 02399-0136 may be a little closer to ``face on'' than
F10214+4724. The effects of differential magnification of the outer
and inner NLRs in SMM 02399-0136 will be much less due to the much 
smaller magnification gradient in the cluster lens, so an infrared 
spectrum of this object should show more normal line ratios. The lack
of a split in the Ly$\alpha$ line could be due to more isotropic
radiation from the outer NLR contributing to the centre of the
profile. 

Hidden or patially hidden star formation in the obscuring tori of 
type-2 AGN may turn out to be
common, and metal-rich outflows from these regions (for which we have evidence
in F10214+4724) may stimulate
starbursts in the remainder of the galaxy (Silk \& Rees 1998). 
It is clear though from studies of F10214+4724 that
disentangling the starburst, AGN
and possibly lensing contributions to the observed 
infrared fluxes from similar objects discovered in future surveys
will prove to be a major challenge. 
This may become possible, for example through very high resolution imaging 
using interferometers working in the submm. If so, selection in the
far-infrared/submm should prove to be much more reliable than UV flux for
tracing the star formation history of the Universe

\section*{Acknowledgements}

We are very grateful to Tom Geballe for obtaining the 
$J$-band spectrum as part of the UKIRT service programme, and to Chris Simpson
for use of his software. We thank the referee, Neil Trentham for a
careful reading of the manuscript and for improving the discussion in
Section 7. We also thank Michel Fioc and Brigitte
Rocca-Volmerange for making their spectral synthesis models freely 
available. The UKIRT is operated by the Joint Astronomy 
Centre on behalf of the U.K. Particle Physics and Astronomy Research Council. 
This research has made use of the NASA/IPAC Extragalactic Database   
which is operated by the Jet Propulsion Laboratory, California Institute  
of Technology, under contract with the National Aeronautics and Space      
Administration.


\begin{thebibliography}{999}

\bibitem{44} Antonucci R., Hurt T., Miller J.S., 1994, ApJ, 430, 210

\bibitem{611} Bohlin R.C., Savage B.D., Drake J.F., 1978, ApJ, 224, 132

\bibitem{10} Broadhurst T., Leh\'{a}r J., 1995, ApJ, 450, L41

\bibitem{612} Bunker A.J., 1996, DPhil thesis, University of Oxford

\bibitem{911} Caganoff S., Antonucci R.R.J., Ford H.C., Kriss G.A.,
Hartig G., Armus L., Evans I.N., Rosenblatt E., Bohlin R.C., Kinney
A.L., 1991, ApJ, 377, L9

\bibitem{613} Calzetti D., Kinney A.L., Storchi-Bergmann T., 1994,
ApJ, 429, 582

\bibitem{34} Cardelli J.A., Clayton G.C., Mathis J.S., 1989, ApJ, 345, 245

\bibitem{614} Condon J.J., 1992, ARA\&A, 30, 575

\bibitem{615} Crenshaw, D. M. 1997, in, Peterson B.M., Cheng F.-Z., 
Wilson A.S., eds,  ASP Conf.\ Proc.\ 113, Emission
Lines in Active Galaxies: New Methods and Techniques. 
ASP, San Fransisco, p.\ 240

\bibitem{91} Davidson K., Netzer H., 1979, Rev.\ Mod.\ Phys., 51, 715 

\bibitem{711} Dey A., van Breugel W., Vacca W.D., Antonucci R., 1997,
ApJ, 490, 698

\bibitem{912} Downes D., Solomon P.M., Radford S.J.E., 1995, ApJ, 453,
L65

\bibitem{50} Eales S.A., Rawlings S., 1993, ApJ, 411, 67

\bibitem{12} Eisenhardt P.R., Armus L., Hogg D.W., Soifer B.T., Neugebauer G.,
Werner M.W., 1996, ApJ, 461, 72

\bibitem{2} Elston, R., McCarthy, P.J., Eisenhardt, P., Dickinson, M.,
Spinrad, H., Januzzi, B.T., Maloney, P., 1994, AJ, 107, 910

\bibitem{73} Ferguson, J.W., Ferland G.J., Pradhan A.K., ApJ, 438, L55

\bibitem{52} Ferland G., Netzer H., 1979, ApJ, 229, 274

\bibitem{564} Fioc M., Rocca-Volmerange B., 1997 A\&A 326, 950

\bibitem{619} Gallego J., Zamorano J., Aragon-Salamanca A., Rego M.,
1995, ApJ, 455, L1

\bibitem{616} George I.M., Turner T.J., Netzer H., Nandra K.,
Mushotzky R.F., Yaqoob T., 1998, ApJS, 114, 73

\bibitem{16} Graham J.R., Liu M.C., 1995, ApJ, 449, L29

\bibitem{11} Goodrich R.W., Miller J.S., Martel A., Cohen M.H., Tran H.D., 
Ogle P.M., Vermeulen R.C., 1996, ApJ, 456, L9

\bibitem{617} Green S.M., Rowan-Robinson M., 1996, MNRAS, 279, 884

\bibitem{620} Guiderdoni B., Bouchet F.R., Puget J.L., Lagache G.,
Hiron E., 1997, Nat, 390, 257

\bibitem{618} Heckman T.M., Gonzalez-Delgardo R., Leitherer C., 
Meurer G.R., Krolick J., Wilson A.S., Koratkar A., Kinney A., 1997,
ApJ, 482, 114

\bibitem{565} Heckman T., 1998, astro-ph/9801155, to appear in the
proceedings of the KNAW colloquium ``The most distant radio galaxies''.

\bibitem{35} Heisler C.A., Lumsden S.L., Bailey J.A., 1996, Nat, 385, 700

\bibitem{33} Hes R., Barthel P.D., Fosbury R.A.E., 1993, Nat, 362, 326

\bibitem{567} Ivison R., Smail I., Le Borgne J.F., Blain A.W., Knieb
J.P., Bezecourt J., Kerr T.H., Davies J.K., 1998, astro-ph/9712161

\bibitem{13} Iwamuro F., Maihara T., Tsukamoto H., Oya S., Hall D.B.N., Cowie
L.L., 1995, PASJ, 47, 265

\bibitem{566} Kinney A.L., Bohlin R.C., Calzetti D., Panagia N., Wyse
R.F.G., 1993, ApJS, 86, 5

\bibitem{46} Koski A.T., 1978, ApJ, 223, 56

\bibitem{41} Kriss G.A., Davidsen A.F., Blair W.P., Ferguson H.C., Long K.S., 
1992, ApJ, 394, L37

\bibitem{17} Kroker H., Genzel R., Krabbe A., Tacconi-Garman L.E., Tecza M., 
Thatte N., Beckwith S.V.W., 1996, ApJ, 463, L55

\bibitem{621} Laor A., Draine B.T., 1993, ApJ, 402, 441

\bibitem{15} Lawrence A., Rigopoulou D., Rowan-Robinson M., McMahon R.G.,
Broadhurst T., Lonsdale C.J., 1994, MNRAS, 266, L41

\bibitem{563} Lowenthal J., Koo D., Guzman R., Gellego J., Phillips
A., Faber S., Vogt N., Illingworth G., 1997, ApJ, 481, 673

\bibitem{622} Madau P., Ferguson H.C., Dickinson M.E., Giavalisco M.,
Steidel C.C., Fruchter A., 1996, MNRAS, 283, 1388

\bibitem{1} Matthews K. et al., 1994, ApJ, 420, L13 (M94)

\bibitem{624} Mathews W.G., Veilleux S., 1989, ApJ, 336, 93

\bibitem{623} Meurer G.R., Heckman T.M., Leitherer C., Kinney A., Robert
C., Garnett D.R., 1195, AJ, 110, 2665

\bibitem{71} Netzer H., 1997, Astrophysics \& Space Science, in press

\bibitem{48} Osterbrock D.E., Fulbright J.P., 1996, PASP, 108, 183

\bibitem{100} Osterbrock D.E., 1989, Astrophysics of Gaseous Nebulae and 
Active Galactic Nuclei. University Science, Mill Valley

\bibitem{36} Radford S.J.E., Downes D., Solomon P.M., Barrett J., Sage L.J., 
1996, AJ, 111, 1021

\bibitem{40} Ramsay S.K., Mountain C.M., Geballe T.R., 1992, MNRAS, 259, 751


\bibitem{950} Rowan-Robinson M., et al., 1997, MNRAS, 289, 490

\bibitem{6} Rowan-Robinson, M., et al., 1993, MNRAS, 261, 513

\bibitem{3} Rowan-Robinson, M., et al., 1991, Nat, 351, 719 

\bibitem{7} Serjeant S., Lacy M., Rawlings S., King L.J., Clements D.L., 
1995, MNRAS, 276, L31 (Paper I)

\bibitem{32} Serjeant S., Rawlings S., Lacy M., McMahon R., Lawrence A., 
Rowan-Robinson M., Mountain M., 1997, astro-ph/9802213, 
MNRAS, in press (Paper II)

\bibitem{991} Silk J., Rees, M.J., 1998, astro-ph/9801013

\bibitem{30} Simpson C.J., Ward M.J., 1996, MNRAS, 282, 797

\bibitem{864} Smail I., Ivison R.J., Blain A.W., 1997, ApJ, 490, L5

\bibitem{951} Smith H.E., Lonsdale C.J., Lonsdale C.J., Diamond P.J.,
1998, ApJ, 493, L17

\bibitem{70} Snijders M.H.J., Netzer H., Boksenberg A., 1986, MNRAS, 222, 549

\bibitem{8} Soifer B.T., Cohen J.G., Armus L., Matthews K., Neugebauer G.,
Oke J.B., 1995, ApJ, 443, L65

\bibitem{14} Trentham N., 1995, MNRAS, 277, 616

\bibitem{712} Verner D.A., Barthel P.D., Tytler D., 1994, A\&AS, 108, 287

\end{thebibliography}
\end{document}